\documentclass[12pt]{iopart}

\usepackage{ragged2e}
\usepackage{graphicx}
\usepackage{soul}
\usepackage{xcolor}
\usepackage[left]{lineno} 

\usepackage{draftwatermark}

\SetWatermarkText{PROOFREAD}
\SetWatermarkScale{3}     
\SetWatermarkLightness{0.9} 

\begin{document}

\title[Topological phase transitions]{Topological Phase Transitions in Superfluids Near Black Hole Horizons}

\author{C R Ghezzi$^{1 *}$ and P S Custodio$^2$}

\address{$^1$ Independent Researcher, 8000 Bahía Blanca, Provincia de Buenos Aires, Argentina
}
\address{$^2$ Universidade Paulista, Campus Marques, Avenida Marquês de São Vicente, 3001, 05036-040 Água Branca,
São Paulo, Brazil 

\textsuperscript{*} Corresponding author: gluon00@yahoo.com
}

\vspace{10pt}
\begin{indented}
\item[]June 2024
\end{indented}

\begin{abstract}
 We investigated a two-dimensional superfluid model immersed in a black hole spacetime and hypothesize that if a black hole collides with a thin superfluid film, it will trigger a topological phase transition within the superfluid, characterized by the production of vortex--antivortex pairs. We adapted the 2D XY model to a curved spacetime and elucidated the topological phase transition in response to variations in the black hole's temperature. Specializing the model to a Schwarzschild--de Sitter black hole, we found a proliferation of vortex--antivortex pairs close to the event and cosmological horizons.
\end{abstract}

\submitto{\PS}

\justifying 

\section{Introduction}
There is increasing interest in simulating quantum fields in curved spacetime backgrounds using condensed matter analog systems. Interestingly, Yang {\it et al.}~\cite{yang} demonstrated that it is possible to simulate a massless scalar or Dirac field in curved spacetime using bosonic hopping models, the free Hubbard model, and the XY model. They also suggested a possible experimental realization in trapped-ion systems. Lamata {\it et al.}~\cite{lamata} proposed a quantum simulator for relativistic quantum mechanics in $2+1$ dimensions using trapped ions. Boada {\it et al.}~\cite{boada} discussed fermions coupled to gravity via ultracold atoms in optical lattices as analog models.

In addition to trapped ions, various physical systems have been proposed for analog gravitational models. For example, analog gravity on graphene and other Dirac-like materials may offer insights into quantum gravity theories (see~\cite{acquaviva} for a review). On more theoretical grounds, Saeed and Husain studied second-order phase transitions of scalar fields in the presence of 4D/5D Schwarzschild and Schwarzschild–AdS black holes using the Ising and Blume–Capel models~\cite{saeed}. Several studies have explored the XY model on curved manifolds by considering classical spin vectors constrained to lie in tangent spaces, where curvature leads to frustration via holonomy effects~\cite{Baek2009}.

The present paper investigates the topological phase transition of a superfluid in Schwarzschild and Schwarzschild–de Sitter black hole spacetimes in $2+1$ dimensions.

Topological phase transitions in laminar (2D) superfluids occur when crossing a critical temperature, resulting in a change in the field's topology~\cite{Kosterlitz}. In material systems, phase transitions are classified by the continuity of derivatives of the Gibbs free energy: discontinuities in first-order derivatives define first-order transitions, while discontinuities in second-order derivatives define second-order transitions. Topological phase transitions, by contrast, are continuous and preserve symmetry~\cite{huang}. The Berezinskii–Kosterlitz–Thouless (BKT) transition involves the unbinding of vortex–antivortex pairs at a critical temperature. This pair formation is analogous to electron–positron pair production in high-intensity electric fields~\cite{Kosterlitz}. Hadzibabic {\it et al.}~\cite{experiment1} observed a BKT-type crossover in a trapped 2D quantum degenerate gas of rubidium atoms.

We study this topological phase transition using the 2D XY model, a two-dimensional spin model in the same universality class as superfluids. Each classical spin in the model can assume any orientation in the plane and interacts with its nearest neighbors. Vacuum fluctuations prevent spontaneous symmetry breaking in two dimensions, as established by the Mermin–Wagner theorem~\cite{mermin}. However, the system may still undergo a continuous topological transition. The numerical methods for the XY model are well developed~\cite{Tobochnik}, and the critical temperature is known with high precision~\cite{Nguyen}.

We propose that if a black hole intersects a cold, laminar superfluid, its Hawking temperature--being inversely proportional to its mass~\cite{hawking1,fabbri}--will induce a topological phase transition. Microscopic black holes, being extremely hot, would have evaporated since their primordial formation if their mass were below $5 \times 10^{14}$ g~\cite{page}. While this model is not intended to describe a realistic astrophysical system, we believe it has theoretical interest due to the increasing interplay between condensed matter analogs of black holes and quantum field theory in curved backgrounds~\cite{yang,visser}.

In this study, we extend the 2D XY model to the spacetime of a black hole and use Monte Carlo simulations to explore the behavior of the scalar field representing the superfluid, modeled as a thin plane at the equator of the black hole.

The outline of the paper is as follows: in Section~\ref{themodel}, we develop the superfluid model in the curved background, define the physical observables measured in the simulations, and describe the numerical algorithm. Section~\ref{metrics} introduces the black hole metrics and their thermodynamic properties. In Section~\ref{results}, we present and discuss our results. Finally, Section~\ref{conclusion} provides our conclusions.

\section{The 2D XY model in curved spacetime} \label{themodel}
A complex scalar field $\psi$ represents the order parameter of the superfluid, as described by the Ginzburg–Landau model, which we generalize to a curved spacetime background. The black hole temperature is treated as an {\it ad hoc} parameter, and we assume that the thin superfluid sheet lies on an equatorial surface, effectively eliminating one degree of freedom.

The simulations depict a scenario in which a probe thin box with thermally conducting walls hovers just above the black hole horizon. The box, considered much smaller than the black hole itself, contains a superfluid in a state of local thermodynamic equilibrium with the surrounding Hawking radiation. In this section, we derive the 2D XY model in curved spacetime in detail, closely paralleling its derivation in flat spacetime~\cite{huang}.

\subsection{Scalar field's energy functional}  \label{secfunctional}
The complex scalar field can be decomposed into its real and imaginary parts: $\psi=(\phi_1+i\,\phi_2)/\sqrt{2}\,,$
while $\psi^*=(\phi_1-i\,\phi_2)/\sqrt{2}\,.$ 
The superfluid has an action given by:
\begin{equation}
S = \int d^{3}x \sqrt{-g} \,\, \mathcal{L} \,,
\end{equation}
with the Lagrangian density given by the following expression:
\begin{equation}
\mathcal{L} =\frac{1}{2} \sum_{i=(1,2)} \left[ g^{\mu\nu} \, \partial_{\mu} \phi_i\, \partial_{\nu} \phi_i - \left(m^2 + \eta\, R\right) {\phi_i}^2 \right] - V(\rho)\,.
\end{equation}
This is the Ginzburg-Landau free energy adapted to curved spacetime.
The first term on the right-hand side of this equation represents the kinetic energy of the field; the middle parentheses contain the bare mass term and the coupling with the scalar curvature of the spacetime; the last term is the potential energy of the field, which could have its usual sombrero form, $V(\rho)=b_1\,(\sqrt{\rho})^2+b_2\,(\sqrt{\rho})^4$, with ${\rho}=\sum_{i=(1,2)} \phi_i^2$. The parameters $b_1$ and $b_2$ can depend on temperature. If the coefficient of the $\rho$ term becomes negative, it would lead to spontaneous breaking of the continuous $U(1)$ symmetry of the Lagrangian in higher dimensions. However, this symmetry breaking does not occur in $2+1$ dimensions due to the Mermin–Wagner theorem.
The rest of the factors have the following meaning: $m$ is the mass of the scalar field,  $R $ is the Ricci scalar for the black hole, $g^{\mu\nu}$ is the inverse metric tensor, and $g$ is its determinant. The indexes $\mu$ and $\nu$ range from $0$ to $2$. With this notation, the complex field is denoted as $\psi(x)=\sqrt{\rho (x)} \,e^{i\,\alpha (x)}$, with $x$ representing space-time coordinates. The derivatives in the kinetic energy term are common partial derivatives since $\phi_i: \phi_i(t, r,\theta) $ is a real scalar field for $i=1, 2$.  
In this paper, we assumed that the field is minimally coupled to gravity, so $\eta=0$. 
The metric for a spherical symmetric black hole is:
\begin{equation} \label{metric}
ds^2 = f(r) \,dt^2 - f(r)^{-1}\,dr^2 - r^2 \,d\phi ^2\,.
\end{equation}
Putting in the metric and integrating over $t$, $r$, and $\theta$ gives the action:
\begin{eqnarray}
S = \int dt\, dr\, d\theta \, r \,\,&\biggl[& \,\frac{1}{2}\,\sum_{i=(1,2)}\, \biggl( f(r)^{-1} (\partial_t \, \phi_i)^2 - f(r) (\partial_r \, \phi_i)^2 \\  \nonumber
&&-\frac{1}{r^2} (\partial_\theta \, \phi_i )^2 -  m^2  {\phi_i}^2 \biggr) -V(\phi) \biggr]\,.
\end{eqnarray}
This action yields the equation of motion for a scalar field on a nonrotating black hole background.
Defining the conjugate momentum of the real field $\phi_i$:
\begin{equation}
\pi_i = \frac{\partial \mathcal{L}}{\partial (\partial_t \phi_i)} = f(r)^{-1} \,\partial_t \phi_i\,,
\end{equation}
the Hamiltonian density is obtained as  the Legendre transform of the Lagrangian: 
\begin{eqnarray}
\mathcal{H} = \sum_{i=(1,2)} (\pi_i \,\partial_t \, \phi_i)  - \mathcal{L} 
&=& \frac{1}{2} \,\sum_{i=(1,2)} \,\biggl( f(r) \, {\pi_i}^2 +   f(r) \, (\partial_r \, \phi_i)^2 \\ \nonumber
&& + \frac{1}{r^2} \,(\partial_\theta \, \phi_i )^2 +   m^2 \, {\phi_i}^2 \biggr) +V(\phi) \,,
\end{eqnarray}
That is, the $T_0^0$ component of the energy-momentum tensor $T_\mu^\nu$. The Hamiltonian is denoted as follows:
\begin{equation} \label{functional}
H = \int dr \,d\theta \, \sqrt{-g}\,\mathcal{H}\,.
\end{equation}
Let us simplify the Hamiltonian density by assuming that the scalar field $\phi$ resides at the minimum of the potential $V(\phi)$. We considered that the potential exhibits a $U(1)$ symmetry, particularly such that $\partial\, V/ \partial \sqrt{\rho} = 0$ at the minimum of the potential.  Considering the stationary case, the conjugate momentum $\pi_i = \partial L/\partial (\partial_t \,\phi_i) = f(r)^{-1} \,\partial_t \,\phi_i \approx 0$. Consequently, the Hamiltonian density simplifies to:
$$ \mathcal{H} = \frac{1}{2} \sum_{i=(1,2)} \biggl(f(r) \,(\partial_r \, \phi_i)^2 + \frac{1}{r^2}\,(\partial_\theta \, \phi_i)^2 \biggr) + V(\rho) \,, $$
Additionally, when $V'' \gg 1$, as observed in the Ginzburg-Landau model, we can safely disregard the Higgs modes of the field, retaining only the Goldstone modes. This simplification is facilitated by assuming the $U(1)$ symmetry in the potential and, thus, in the Lagrangian.
By neglecting the Higgs modes, only the angular coordinate $\alpha$ remains as the primary descriptor for the Goldstone modes. Thus, we can express the field as $\psi=\sqrt{\rho_0}\, e^{i \,\alpha}$, where $\rho$ remains constant at a specific value ($\rho=\rho_0$), as explained above. With these considerations in mind, the Hamiltonian density assumes a more concise form:
\begin{equation} \label{hamilden}
 \mathcal{H} = \frac{1}{2} \bigg( f(r)\, \rho_0 \,(\partial_r \, \alpha)^2 + \frac{1}{r^2}\, \rho_0\,(\partial_\theta \, \alpha)^2 \bigg) + V(\rho_0)  \,.
\end{equation}
In this context, we omit the partial derivatives of $\rho$ to align with the underlying assumptions, expressing the potential as a function of the constant $\rho_0$. In Minkowski spacetime, the metric coefficient is $f(r)=1$, effectively recovering the Hamiltonian density for the 2D XY model from the equation above. The energy functional commonly employed for Minkowski spacetime is:
\begin{equation} \label{minkowski}
H = \int dx \,dy \,  \biggl[\frac{1}{2} \rho_0 \,(\nabla \alpha)^2+V(\rho_0) \biggr]\,.
\end{equation}
In the case of the black hole spacetime, we can apply the principles of lattice field theory to the energy functional, as presented in Eq.~(\ref{functional}), featuring the Hamiltonian density outlined in Eq.~(\ref{hamilden}). Before proceeding with the discretization process, several substitutions within Eq.~(\ref{hamilden}) are necessary: we rename $\partial_r$ as $\partial_x$, $r^{-1} \partial_\theta$ transforms into $\partial_y$, $dr$ is replaced by $dx$, and substituted $r\,d\theta$ with $dy$. 
{{The difference in $r\,d\theta$ between the top and bottom of the box is second order in the differentials and therefore negligible, since $l_{box} \ll r_H$, where $l_{box}$ denotes the finite linear dimension of the box.}} Therefore, we have neglected this difference. In essence, this approach results in the energy functional for the 2D XY model in a black hole spacetime (modulo a constant value), which we summarized as:
\begin{equation} \label{functional2}
H = \frac{1}{2}\,\rho_0\, \int dx \,dy \, \biggl[f(r)\,(\nabla_x \,\alpha)^2+ (\nabla_y \, \alpha)^2\biggr]\,.
\end{equation}
A direct comparison between the energy functionals of the 2D XY model in Minkowski spacetime, as expressed in 
Eq.~(\ref{minkowski}), and the black hole spacetime, as shown in Eq.~(\ref{functional2}), provides insight into the effects of the black hole background on the superfluid dynamics, i.e., the superfluid is now anisotropic due to the curved spacetime. Essentially, the radial interaction between spin neighbors is redshifted, while the tangential interaction is not. 

For the subsequent discretization step, we must make the following substitutions: $\int dx \,dy$ by $a^2 \sum_{\langle x y \rangle}$, $\partial_y \alpha$ is transformed into $a^{-1} (\alpha _{y+1}-\alpha _y)$, and $\partial_x \alpha$ is rewritten as $a^{-1} f(r) (\alpha _{x+1}-\alpha _x)$.
In this notation, it is essential to understand that $(\alpha _{y+1}-\alpha _{y})$ is equivalent to $(\alpha _{x, y+1}-\alpha _{x, y})$ and so forth, with the fixed subindex omitted for clarity. The symbol $a$ denotes the lattice spacing. Consequently, the energy functional in its discrete form takes shape as follows:
\begin{equation} \label{discrete}
H=\frac{1}{2}\, \rho_0 \, a^2 \,\sum_{\langle x y \rangle} \biggl[ f(r)\, a^{-2} \,(\alpha_{x+1}-\alpha_x)^2 +a^{-2} 
(\alpha_{y+1}-\alpha_y)^2 \biggr] \,.
\end{equation}
 We can see that the lattice spacing cancels out and that the strength of the interaction between neighbors depends exclusively on the squared order parameter modulus $\rho_0$. It is possible to rewrite this expression to more closely resemble the flat 2D XY model by replacing $\cos(\alpha_{x+1}-\alpha_x) \sim 1-\frac{1}{2} (\alpha_{x+1}-\alpha_x)^2$, the Eq.~(\ref{discrete}) becomes:
\begin{equation} \label{discrete2}
H=- \rho_0 \,\sum_{\langle x y \rangle} \biggl[ f(r)  \cos(\alpha_{x+1}-\alpha_x) + 
\cos(\alpha_{y+1}-\alpha_y) \biggr] \,.
\end{equation}
{{
This is the final form of the energy functional for the 2D XY model in a black hole spacetime. In the double sum of Eq.~(\ref{discrete2}), $x$ and $y$ range from $1$ to $N$. 

As the energy functional does not explicitly depend on the physical lattice dimensions, we may identify the lattice spacing with the superfluid coherence length $\xi$, which is inversely proportional to the square root of the boson mass~\cite{huang}. In this way, the physical size of the simulation box becomes $l_{box} \approx N\, \xi$, and the condition $l_{box} \ll r_H$ ensures that the box remains small compared to the black hole horizon. At the same time, $N$ must be large enough to accommodate multiple vortex–antivortex pairs. Our results are independent of this scaling and can be generalized to black holes of arbitrary size, provided this condition is satisfied.

Equation~(\ref{discrete2}) can be further simplified by denoting each site by a single index $i$ corresponding to the spatial coordinates $(x,y)$, and by expressing the model in terms of 2D unit vectors, denoted as \( \mathbf{s}_i = (\cos \alpha_i, \sin \alpha_i) \). Here, $i+1$ denotes the site $(x+1, y)$ and $j+1$ the site $(x, y+1)$. 

Using this notation and standard trigonometric identities, the cosine functions in Eq.~(\ref{discrete2}) can be rewritten as Euclidean scalar products. The first cosine is the scalar product $(\mathbf{s}_i, \mathbf{s}_{i+1}) = \cos(\alpha_{x+1} - \alpha_x)$ with fixed $y$, while the second cosine is $(\mathbf{s}_j, \mathbf{s}_{j+1}) = \cos(\alpha_{y+1} - \alpha_y)$ with fixed $x$.

We emphasize that these vectors represent the local phase of the field at site $(x, y)$, and should not be interpreted as classical spin vectors in the differential geometric sense (i.e., as tangent vectors on a curved spacetime manifold). Although this terminology is commonly used in flat-space studies of the XY model, here it refers strictly to abstract internal degrees of freedom encoding angular information. For consistency with the established literature on the XY model, we will continue to refer to these 2D unit vectors as ``spins.''

We now define a modified scalar product between neighboring spins as follows. For neighbors along the radial ($x$) direction:
$
\mathbf{s}_i \cdot \mathbf{s}_{i+1} = (\mathbf{s}_i, \mathbf{s}_{i+1})\, f(r)\,,
$
and for neighbors along the transverse (\(y\)) direction:
$
\mathbf{s}_j \cdot \mathbf{s}_{j+1} = (\mathbf{s}_j, \mathbf{s}_{j+1})\,.
$

Thus, Eq.~(\ref{discrete2}) reduces to:
$$
H = - J\, \sum_{\langle ab \rangle} \mathbf{s}_a \cdot \mathbf{s}_b\,,
$$
where the sum extends over all pairs of neighboring sites $\langle ab \rangle$ in the radial and angular directions. We set $J = \rho_0$ for the coupling strength.
}}

The expression above shows that the resulting energy functional is $U(1)$ invariant and preserves the symmetry of the Lagrangian, although it is spatially anisotropic due to the curved background. We note that it reduces to the standard 2D XY model in the flat case, where $f(r)=1$.

Thus, the derivation of the energy functional for the 2D XY model in a curved background is complete.

\subsection{Measurements}
We want to emphasize the relation between fields and statistical mechanics. 
In thermal field theory in $d$-dimensions, the Euclideanized quantum path integral of the vacuum correlation function, with a compactified temporal direction of length $\beta=1/T$ (see \cite{huang}, pag. 274), is:
\begin{equation}
Z=\int {\cal D}\phi \,e^{-S(\phi)}\,.
\end{equation}
There is a well-known connection between the path integral and the ensemble average in statistical mechanics established by identifying the Euclidean action with the energy functional \cite{huang}: $S(\phi)= \beta \, H(\phi)$.
In the limit $\beta \rightarrow 0$ and the space volume tending to infinity, the path integral above reduces to the partition function of a classical field in $d-1$ dimensions. On the other hand, when $\beta \rightarrow \infty$, the path integral corresponds to a quantum field theory in $d$ dimensions with zero temperature.
Thus, we can write the partition function more appropriately:
\begin{equation}
Z=\Pi_i \int_0^{2 \pi} d \alpha_i \,e^{-\beta \,H(\alpha)}\,.
\end{equation}
The subindex $i$ indicates that the product is over all the spatial points that support the field. Since the lattice is discrete, a numerable set covers the scalar field.

Some relevant observables of the field are the correlation function, the specific heat, the magnetic susceptibility, the stiffness, and the number of vortices.
The correlation function is given by:
\begin{equation}
G(r, T)=\left < \mathbf{s_i \,.\,  s_j} \right>_T-\left < \mathbf{s_i } \right>_T\, \left < \mathbf{s_j} \right>_T \,,
\end{equation}
where the brackets with a subscript $T$ indicate a thermal ensemble of the enclosed physical observables.
The correlation function signals the topological phase transition since at temperatures above critical, it decays exponentially, and at lower than critical temperatures, it has a quasi-long range order, i.e., following a power law decay. However, it is difficult to determine the transition temperature using the correlation function \cite{Kosterlitz}. Therefore, we will not analyze it.
 The specific heat is defined as:
\begin{equation} \label{specific_heat}
c_v=\frac{\partial \left <E \right>}{\partial T}=\frac{k_B\, \beta}{N} \,\biggl( \left < {E^2} \right>_T-\left < {E} \right>^2_T \biggr)\,,
\end{equation}
where we restored the units, so $\beta=(k_B T/J)^{-1}$, and the number of spins $N \rightarrow \infty$.
Another important observable is the number of vortices and antivortices as a function of temperature. Consider the phase angle $\alpha: \alpha(x)$ as a function of the coordinates $x$ of a point on the lattice. Angle $\alpha$ can change modulo $2 \pi$, thus the integral on a closed path $C$ can vary only an integer number of times $2 \pi$:
\begin{equation} \label{charges}
\oint_C d\mathbf{s}\,.\,\nabla \alpha= 2\, \pi \,n  \,\,\,\,\,\,\,\,\,\, (n=0, \pm 1, \pm 2, ...)
\end{equation}

A closed path $C$ cannot contract to a point if $n$ is different from zero; the field is multiply connected in this case. The singular points in a multiply connected field are referred to as ``charges" due to an analogy between electric charges and vortices \cite{Kosterlitz}, with the added feature of quantized integrals.  The elements of the field homotopy group characterize the charges. Conventionally, charges with $n=+1$ are identified as vortices, while those with $n=-1$ are labeled antivortices. Charges with $|n|>1$ are statistically improbable in the current model.

When there is at least one vortex, the field becomes multiply connected, with vortices encapsulating normal fluid at their cores. The desired observables are the total number of vortices $N_+$ and antivortices $N_-$ as a function of temperature. Conservation of total charge occurs as the size of the lattice approaches infinity, ensuring that $N_+-N_-=0$. The BKT phenomenon signifies that pair production becomes increasingly prominent with increasing temperatures. Initially, vortex--antivortex pairs form closely bonded configurations and gradually separate as the temperature climbs. Beyond the critical temperature, these pairs become ionized, allowing vortices and antivortices to move freely around the lattice \cite{Kosterlitz}.

The other essential physical quantity is the spin stiffness, which measures the response of the spins to a modulation of the order parameter \cite{Sandvik}. In other words, the stiffness gives the rigidity of the lattice under a twist of the spins at the boundaries. We define spin stiffness as $\rho_s=N^{-1}\,\partial^2 F(\gamma)/\partial \gamma^2$, where $F(\gamma)$ is the free energy as a function of the spin twist $\gamma$ at the boundary. In a thermal ensemble, we calculate the spin stiffness as follows:
\begin{equation}
\rho_s=\frac{1}{N} \biggl( \langle H_x \rangle-\langle I_x^2 \rangle \bigg)\,,
\end{equation}
where: $H(\gamma)=-J\,\sum_{\langle i,j \rangle_x} \cos(\theta_i-\theta_j+\gamma)-J\,\sum_{\langle i,j\rangle_y} \cos(\theta_i-\theta_j)$ and $H_x=\partial H(\gamma)/\partial \gamma \,|_{\gamma=0}$, and the total spin ``current" in the $x$ direction is $I_x=J\,\sum_{\langle i,j \rangle_x} \sin(\theta_i-\theta_j)$ (see Ref. \cite{Sandvik}, pag. 49, for details).
The stiffness allows us to measure the transition temperature because the equations of the Kosterlitz renormalization group predict a jump in stiffness from the value $2 \,T_c/\pi$ to zero at the transition temperature \cite{minnhagen}. Therefore, the intersection of the line $2\, T/\pi$ with the stiffness graph as a function of temperature, when the size of the lattice tends to infinity, gives the critical temperature for the topological phase transition.

The following subsection outlines the specific details of the numeric algorithm used to simulate the model described in Subsection \ref{secfunctional}. These simulations enable us to perform the measurements outlined in the present subsection.

\subsection{Monte Carlo implementation}
The Monte Carlo simulation evaluates the change in energy after each spin's random angle rotation. The algorithm computes the energy functional derived in Subsection \ref{secfunctional} for a particular field configuration.
A Monte Carlo (MC) sweep randomly flips each spin of the lattice and computes the change in energy each time. The criterion for selecting a particular spin flip is given by the Metropolis algorithm, as described in several textbooks (see, for example, \cite{huang, Kooning}).
As in the flat 2D XY model, each spin can adopt an angle in the interval $[0, 2 \pi)$.
The simulation can have a ``cold start," where all the spins are initially aligned, or a ``hot start," with all the spins at random orientations. The random spin flips help the simulation avoid getting stuck in local energy minima and allow for a thorough exploration of the system's entire phase space. We do not address the interaction of each spin with the curved spacetime.

In our implementation, the thermalization phase consists of $100$ MC sweeps, followed by approximately $3000$ groups of $100$ MC sweeps at each temperature. We take the measurements at the end of each $100$ MC sweep cycle. This approach ensures a large ensemble of non-correlated measurements. After conducting several experiments, we found this setup optimal for performing the desired numerical investigations and for reproducing the simulations of the 2D XY model \cite{Tobochnik,ghezzi}.

We performed the charge measurement by approaching the integral of Eq.~(\ref{charges}) with a discrete sum over the elements of a plaquette, i.e., summing over the phase angles of the four vertices of a plaquette modulo $2 \pi$; This sum must be $\pm 1$ \cite{Tobochnik}. 

 We performed simulations on a $20 \times 20$ grid, with temperature increments of $0.02\, K$. Previous studies on the 2D XY model temperature allowed us to implement a small grid while achieving accurate results with bounded numerical error \cite{Tobochnik,Nguyen,ghezzi,minnhagen}. The size of the superfluid box’s sides is $(1/10)^{th}$ of the event horizon radius $R_H$, making the application of periodic boundary conditions appropriate.

\begin{figure}  
    \centering
    {\includegraphics[width=85mm]{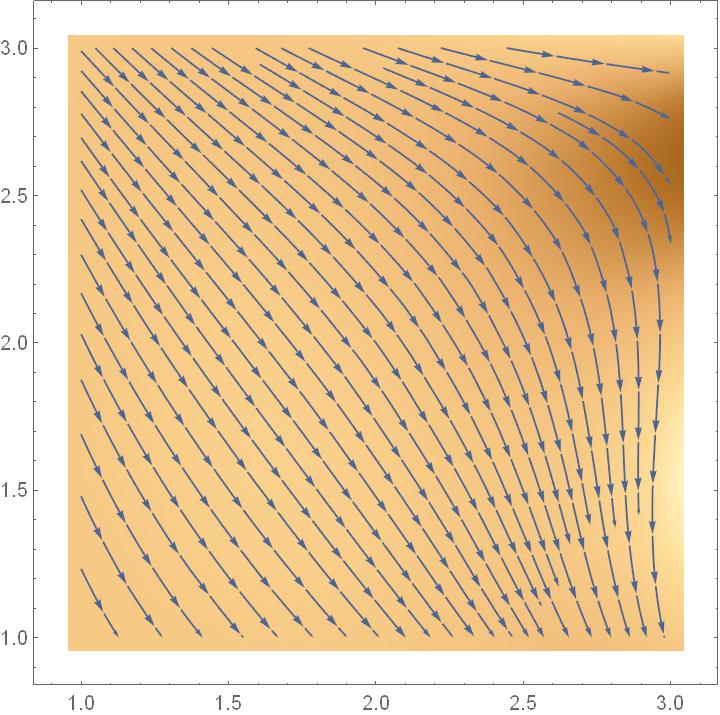}} 
    {\includegraphics[width=85mm]{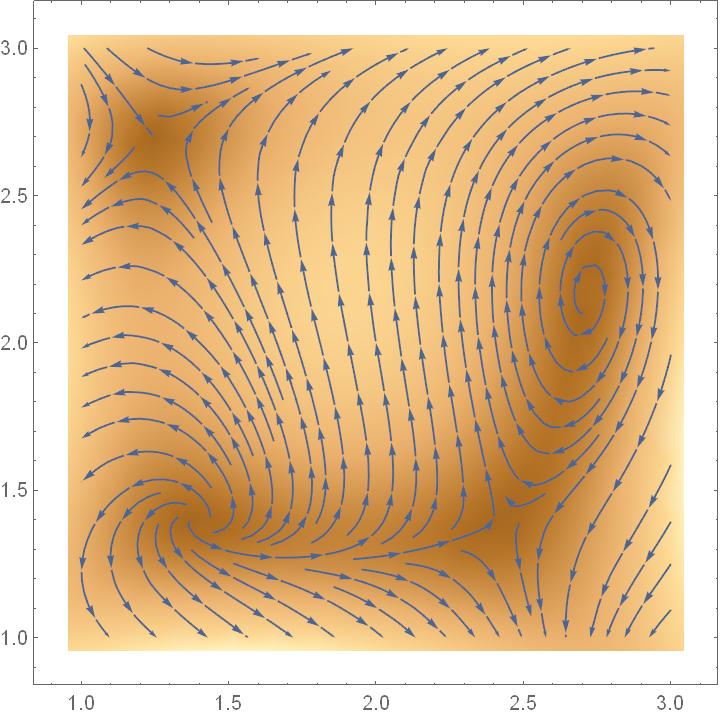}} 
    \caption{{{Vortices and antivortices in a window of $(1/6)^{th}$ the lattice size, for the flat case. The top image shows the streamline-density plot at a temperature below the critical point, where no pairs are present. The bottom graph displays two vortex--antivortex pairs at a temperature above the critical point.}}}
    \label{fig:vortex}
\end{figure}
\begin{figure}
\centering
{\includegraphics[width=85mm]{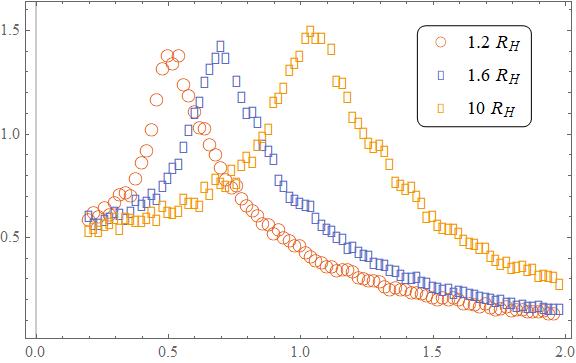}}
{\includegraphics[width=85mm]{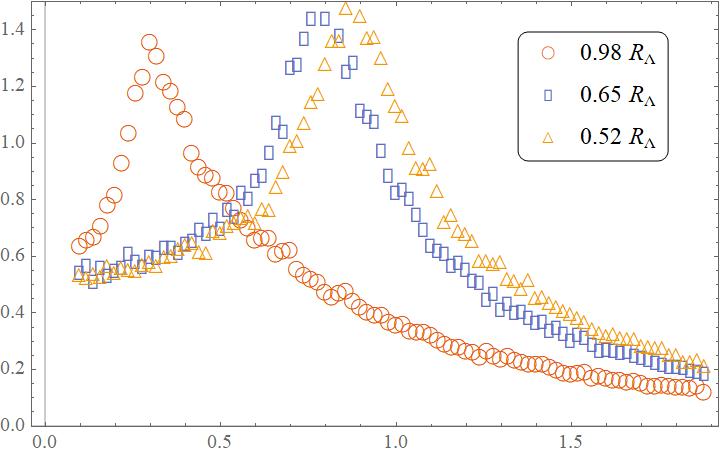}}
  \caption{Specific heat as a function of black hole temperature, with each data set representing three distinct distances from the black hole's horizon, as indicated in the legend. The top graph illustrates the specific heat for a superfluid near the event horizon of a Schwarzschild black hole, while the bottom graph shows the specific heat at a distance from the cosmological horizon of a Schwarzschild--de Sitter black hole.
  }
  
  \label{fig:calor}
\end{figure}

\section{Application to a black hole metric}  \label{metrics}
We consider the superfluid sheet immersed in the gravitational field of a Schwarzschild black hole, and we also explored the case of a Schwarzschild--de Sitter black hole. In both cases, the superfluid is in an equatorial plane of the black hole, so one angular coordinate can be suppressed, and the problem is effectively two-dimensional. Therefore, the system consists of a superfluid plane plus a spherical symmetric black hole. We neglected the backreaction problem. The following subsections will describe the essential properties of these black holes. 

\subsection{Schwarzschild black hole}
The line element for a non-rotating black hole is given by 
Eq.~(\ref{metric}).
Throughout this paper, the employed units of measure are geometric, thus: $G=1$ and $c=1$.
Note that the metric is static and $g_{t t}=-g_{r r}^{-1}=f(r)$. For a 4D Schwarzschild black hole,  $f(r)=1- 2 M/r$. The black hole horizon is at $r=2 M$, and according to the four laws of black hole thermodynamics, it has mass, temperature, and entropy \cite{bardeen}. The space is asymptotically flat. 
The Schwarzschild's black hole asymptotic temperature is
$T_{h_{}}=({8 \,\pi\, M})^{-1}\,.$
{{Near the black hole horizon, the Hawking radiation of a Schwarzschild black hole appears approximately Planckian. The corresponding blueshifted (fiducial) temperature near the horizon is given by~\cite{susskind}:
$T'_{h_{}}=1/{4 \pi \sqrt{2\, M\, r\, (1-{2 M}/{r})}}\,.$
At spatial infinity, this expression reduces to the Hawking temperature $T_h$ when accounting for the gravitational redshift factor. }}

\subsection{Schwarzschild--de Sitter black hole}
{{The metric for the Schwarzschild--de Sitter black hole is defined by the line element as presented in Eq.~(\ref{metric}), where $f(r)=1- 2 \,M/r-\Lambda \,r^2/3$. }} This black hole has two horizons, occurring when $\Lambda>0$ and $9\, \Lambda\, M^2<1$. These horizons manifest as the two positive roots of the metric function $f(r)$, representing the black hole horizon $R_H$ and the cosmological horizon $R_{\Lambda}$.
If the mass of the black hole is zero, $M=0$, the solution reduces to the de Sitter spacetime, characterized by a cosmological horizon located at $R_{\Lambda}=3^{1/2} \,\Lambda^{-1/2}$. Its scalar curvature is positive. The Schwarzschild--de Sitter black hole is asymptotically de Sitter as $r \rightarrow \infty$. It has an area denoted by $A= 4 \,\pi \,R_{\Lambda}^2$, associated with entropy. Gibbons and Hawking \cite{gibbons} demonstrated that this horizon exhibits a temperature of $T_{\Lambda}=(2\, \pi\, R_{\Lambda})^{-1}$.
Therefore, in this spacetime, the observers measure the Hawking radiation at temperature $T_h$ coming from the black hole event horizon and are surrounded by a cosmological horizon emitting at temperature $T_{\Lambda}$.

\begin{figure}  
    \centering
    {\includegraphics[width=85mm]{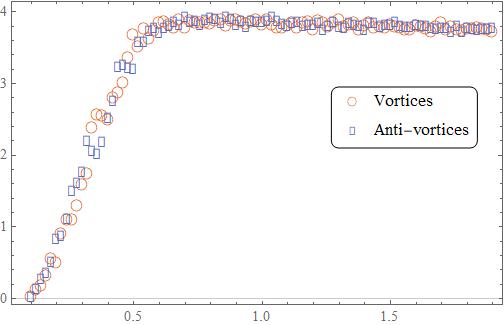}} 
    {\includegraphics[width=85mm]{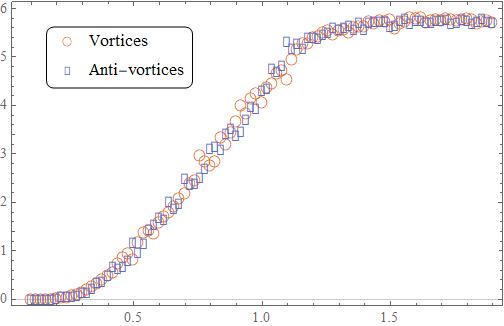}} 
    \caption{Number of vortices as a function of temperature. Top: at a radius $r=1.2\,R_H$; Bottom: at $r=10\,R_H$.}
    \label{fig:number_of_vortex}
\end{figure}
\begin{figure}  
    \centering
    {\includegraphics[width=85mm]{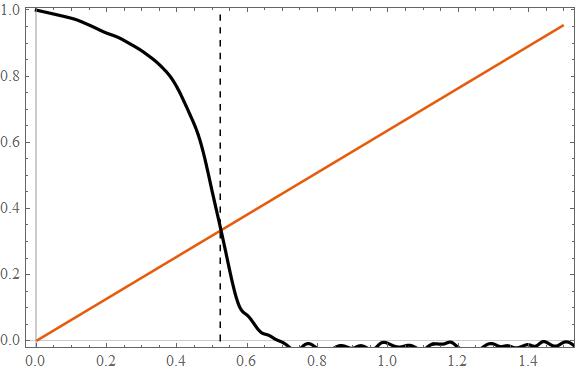}} 
    {\includegraphics[width=85mm]{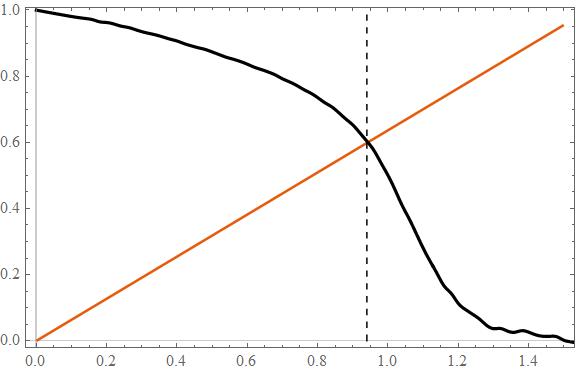}} 
    \caption{Spin stiffness. Top: at $r=1.2\,R_H$ from the event horizon; Bottom: farther from the horizon, at $r=10\,R_H$. The data points are connected with a line to guide the eye.}
    \label{fig:spin}
\end{figure}

\section{Results}   \label{results}
Figure \ref{fig:vortex} illustrates the distinctive behavior of the superfluid topological phase transition. The top figure shows the superfluid in flat spacetime, below the critical temperature, zoomed in at a small window $1/6\,^{th}$ of the simulation lattice. The superfluid does not exhibit any topological defects. The bottom image shows the same window at a higher than critical temperature, where two vortex--antivortex pairs are visible.

Figure \ref{fig:calor} (top) depicts the specific heat as a function of the temperature for the superfluid at three different distances from the event horizon: $1.2 \,R_H$, $1.6 \,R_H$, and $10\, R_H$. We obtained these results measuring the specific heat as indicated in Eq.~(\ref{specific_heat}). The maximum of the curves is closely related to the critical temperature for a topological phase transition \cite{Kosterlitz}. The figure shows that as the field approaches the horizon, the critical temperature for the phase transition decreases, tending toward zero at the event horizon, $R_H$.
For any given distance from the event horizon, a critical BH mass exists that induces a topological phase transition in the superfluid.
For example, for Schwarschild black holes, at a given distance from the event horizon, say $1.2\, R_H$, knowing the critical temperature is $T_{c}\sim 0.53\, K$, the corresponding critical mass can be determined using $M_c=(8 \, \pi \, T_{c})^{-1}$. This result means that black holes with a mass lower than $M_c$ will produce vortex--antivortex pairs at $r=1.2\,R_H$. However, black holes with a mass greater than critical will not be hot enough to produce vortex--antivortex pairs at that distance.
For Schwarschild black holes, the critical black hole mass gets smaller (hotter BH) as the superfluid box is farther out from the event horizon. However, there is a minimum mass as we approach a flat spacetime farther out enough. As shown in Figure \ref{fig:calor} (top),
at a distance of $10\, R_H$, the spacetime does not notably modify the superfluid's thermodynamic properties concerning the flat case. Therefore, at $10\,R_H$, the heat capacity is indistinguishable from the flat spacetime values. The specific heat curve maintains the characteristic shape of the rescaled curve, as depicted in Fig. \ref{fig:calor} (top). A similar trend is evident in Fig. \ref{fig:number_of_vortex}, which illustrates the variation in the count of vortices and antivortices with temperature; the saturation of the curve shifts to lower temperature regimes in proximity to the event horizon (Fig. \ref{fig:number_of_vortex}, top).

The third observable we investigated is the spin stiffness, which we used to determine the topological phase transition temperature. 
Weber and Minnhagen~\cite{minnhagen} derived an interpolation formula for the spin stiffness based on finite-size scaling, allowing accurate estimation of the critical temperature even for relatively small lattice sizes (as small as $\sim 3 \times 3$). Their method accounts for finite-size effects by identifying the temperature at which the stiffness curve intersects the line $2T/\pi$, providing a reliable estimate of $T_c$ in the thermodynamic limit.

In our study, rather than focusing on obtaining precise numerical values, our goal was to highlight the novelty of the physical effects. We present the spin stiffness as a function of temperature in Fig.~\ref{fig:spin}.
On the top, we depict the stiffness of a superfluid near the black hole horizon, while at the bottom, it illustrates the stiffness farther from the horizon. The vertical dashed line intersects the graph at the point where it crosses the line $2 \,T/\pi$, pinpointing the critical temperature on the horizontal axis. Utilizing the stiffness, we calculated a critical temperature of $T_c = 0.93\, K$ for the superfluid at $r=10\,R_H$ (far from the black hole) and a critical temperature of $T_c =0.52\, K$ for the superfluid at $r=1.2\,R_H$ (near the black hole). However, it's worth noting that the critical temperature in flat spacetime tends to be overestimated in finite-size lattices, as suggested by the Weber-Minnhagen study \cite{minnhagen}. In our investigation, $T_c$ is approximately $6 \,\%$ higher with a $20 \times 20$ lattice. The most accurate value of the critical temperature in flat spacetime is $T_c=0.887\,K$, with an error in the tenth percent range \cite{Tobochnik,Nguyen,minnhagen}.

In simulations involving a Schwarzschild–de Sitter black hole, we adjust the cosmological constant to set the cosmological horizon at a finite distance from the black hole horizon. In geometric units with mass \( M = 1 \), the black hole horizon lies at \( R_H = 2 \), and the cosmological horizon at \( R_{\Lambda} = 30.6 \), which is slightly smaller than the pure de Sitter value \( (3/\Lambda)^{-1/2} = 31.62 \). These values correspond to a black hole of approximately \( 7.5 \times 10^{-4} \, M_{\odot} \) and a Schwarzschild radius of about 22 km.
For comparison, vortex core sizes in YBCO superconducting thin films have been measured at around \( 6\,\mu{\rm m} \)~\cite{wells}. This supports the validity of approximations used in our model, such as treating the superfluid as thin and the simulation box as small relative to the black hole.
A bosonic field composed of ultralight particles (e.g., \(10^{-23}\)--\(10^{-20}\,{\rm eV}\)) could, in principle, support vortices on much larger scales, although their quantitative behavior remains unexplored.

It is worth noting that the results obtained are independent of the black hole mass and are applicable across a wide range of mass values.
In this spacetime, we observe results similar to those of the Schwarzschild scenario near the black hole horizon. However, when situated at an intermediate distance between the black hole and cosmological horizons, we notice a slightly lower critical temperature concerning the flat case and the Schwarzschild far field.
As depicted in Fig. \ref{fig:calor} (bottom), at a distance of  $r=0.52 \, R_{\Lambda}$, the peak of the specific heat curve occurs at $0.85\,K$, whereas in the asymptotic region of the Schwarzschild black hole, the peak is at $1.05 \, K$, illustrated on the top side of the figure. 
These results are due, of course, to the presence of the cosmological horizon. 
Inspecting the superfluid closer to the cosmological horizon, we observed that the critical temperature tends to zero as it approaches the cosmological horizon, similar to the behavior near the black hole horizon. As depicted in Fig. \ref{fig:calor} (bottom), at a distance of $r=0.65\,R_{\Lambda}$, the peak of the specific heat shifts to $0.76\,K$, and at  $r=0.98 \, R_{\Lambda}$, the peak termperature  reduces to $0.3\,K$. These results imply a reduction of the critical temperature close to both horizons. 

One notable consequence of the spacetime metric is the induction of anisotropy within the superfluid. This anisotropy in the particle interaction strength arises from the metric coefficient, $f$, which acts as a multiplier in the radial interaction (see Eq.~\ref{discrete2}). As $f$ approaches unity farther away from the horizon, the model gradually transitions towards isotropy, resembling the scenario in flat spacetime. In the Schwarzschild--de Sitter spacetime, which is not asymptotically flat, the flat spacetime behavior is effectively recovered at an intermediate distance between the black hole and cosmological horizons, assuming that these horizons are sufficiently far apart. {{We note that our Hamiltonian is formally equivalent to the anisotropic XY model studied on flat lattices with direction-dependent couplings~\cite{Spisak}. In our case, however, the anisotropy arises naturally from the curved background geometry.}}

Therefore, a topological phase transition occurs on a superfluid when embedded in a curved spacetime. The larger the mass of the black hole, the closer to the horizon we have to look for loose vortex pairs. The extent of the halo of vortex--antivortex pairs surrounding the black hole will be more significant with higher black hole temperatures.
The cosmological horizon also has a halo of pairs on the same side as the observer. In other words, the formed pairs are interior to the cosmological horizon. 
 Altered thermodynamic properties of the superfluid arise from the geometry of the black hole spacetime.
{The critical temperature tends to zero at both horizons and the formation of vortex--antivortex pairs are analogous to the creation of electron–positron pairs in an external electrostatic field. This phenomenon, in turn, has been noted as analogous to Hawking radiation (see~\cite{frolov}, p.~349), with an important distinction: vortex--antivortex pairs are formed outside the black hole's event horizon, whereas in Hawking radiation, energy conservation requires one member of the pair to fall into the black hole~\cite{ruffini}.}
 
Although we presented the results of a toy model, there could be scenarios where it holds relevance. For instance, if dark matter comprises ultralight bosons, Huang {\it et al.} \cite{huang2}  proposed the existence of a vortex lattice surrounding supermassive black holes using a different mechanism involving angular momentum quantization and frame dragging due to the rotating black hole. When considering the backreaction of the superfluid, it is necessary to employ the exact solution for a disk combined with a black hole \cite{letelier}, which is beyond the scope of this paper.
Furthermore, as the formation of vortex--antivortex pairs mirrors electron--positron pair formation, another intriguing application could be its study related to the collapse of charged or polarized compact objects \cite{ghezzi2}. 

\section{Conclusion} \label{conclusion}
In this paper, we investigated a model of a superfluid influenced by the gravitational field of a black hole, employing an adapted 2D XY model. Our simulations focused on the superfluid within a region much smaller than the radius of the black hole, revealing significant modifications in its thermodynamic properties based on its proximity to either the black hole horizon or the cosmological horizon.
Our model simplified the radiation transport through the superfluid, assuming a thermal reservoir at the Hawking temperature. As the critical temperature for the topological phase transition approaches zero near the event or cosmological horizon, a halo of vortices and antivortices forms around these horizons. Notably, the hotter the black hole or the smaller its mass, the broader the region surrounding the black hole hosting vortex--antivortex pairs.
In the case of Schwarzschild--de Sitter black holes, the temperature of the cosmological horizon increases with the cosmological constant. Consequently, the production of the vortex--antivortex pairs is interior to the spherical cosmological horizon, and the extent of the halo expands with the increasing temperature of the cosmological horizon.
{The production of vortices and antivortices near the event horizon is analogous to the formation of electron--positron pairs in a high-intensity electrostatic field, and bears some resemblance to the Hawking radiation phenomenon, although it is not exactly equivalent. }

We found that the event (or cosmological) horizon consistently produces vortex--antivortex pairs, and that their density may decrease with distance from the horizon, following the variation in the critical temperature due to the anisotropy induced by the metric.


\medskip

\begin{thebibliography}{99}

\bibitem{yang} Yang R Q, Liu  H, Zhu S, Luo  L and  Cai  R G, 2020 ``Simulating quantum field theory in curved spacetime with quantum many-body systems'' {\it Phys. Rev. Research} {\bf 2} 2 023107

\bibitem{lamata} Lamata L,  Casanova J,  Gerritsma R,  Roos C F,  García-Ripoll J J and  Solano E 2011 {\it New J. Phys.} {\bf 13} 095003

\bibitem{boada} Boada O, Celi A, Latorre J I and  Lewenstein M 2011 {\it New J. Phys.} {\bf 13} 035002

\bibitem{acquaviva} Acquaviva G, Iorio A, Pais P and Smaldone L 2022 ``Hunting Quantum Gravity with Analogs: the case of graphene'' {\it Universe}  {\bf 8} 9 455

\bibitem{saeed} Saeed M A and V Husain 2024  ``Ising-like models on Euclidean black holes" {\it Class. Quant. Grav.} {\bf 41} 015002, arXiv:2306.08547

\bibitem{Baek2009} Baek S K, Hiroyuki S and Beom J K 2009 ``Curvature-induced frustration in the XY model on hyperbolic surfaces" {\it Physical Review E} {\bf 79} 060106
 
\bibitem{Kosterlitz} Kosterlitz J M  1974 ``The critical properties of the two-dimensional XY model" {\it J. Phys. C: Solid State Phys.} {\bf 7} 6 1046
\newline \noindent
  Kosterlitz J M 2016  ``Kosterlitz-Thouless physics: A review of key issues" {\it Rep. Prog. Phys.} {\bf 79} 2 026001  

\bibitem{huang}  Huang K 1998 {\it Quantum Field Theory. From Operators to Path Integrals} (John Wiley \& Sons)

\bibitem{experiment1} Hadzibabic  Z, Krüger P, Cheneau M, Battelier B and Dalibard J 2006 ``Berezinskii–Kosterlitz–Thouless crossover in a trapped atomic gas'' {\it Nature} {\bf 441} 7097 1118-1121

\bibitem{mermin} Mermin N D and  Wagner  H  1996 ``Absence of ferromagnetism or
antiferromagnetism in one- or two-dimensional isotropic Heisenberg models'' {\it Phys. Rev. Lett.} {\bf 17} 1133–1136   

\bibitem{Tobochnik}  Tobochnik J and  Chester  G V 1979 ``Monte Carlo study of the planar spin model" {\it Phys. Rev. B}  {\bf 20} 9  3761
\newline \noindent
Miyashita  S, et al. 1978 ``Monte Carlo simulation and static and dynamic critical behavior of the plane rotator model," {\it Progress of Theoretical Physics} {\bf 60} 6  1669-1685
\newline
Kenna R 2005 ``The XY model and the Brezinski-Kosterlitz-Thouless phase transitions" arxiv: cond-mat/0512356

\bibitem{Nguyen} Nguyen P H and Boninsegni M 2021 ``Superfluid transition and specific heat of the 2D XY model: Monte Carlo simulation" {\it Applied Sciences} {\bf 11}  11 4931

\bibitem{hawking1} Hawking S W  1975 ``Particle creation by black holes" {\it Commun. Math. Phys.} {\bf 43} 3 199-220

\bibitem{fabbri} Fabbri A and  Navarro-Salas J 2005 {\it Modeling black hole evaporation} (World Scientific)  

\bibitem{page} Page D N  1976 ``Particle emission rates from a black hole: Massless particles from an uncharged, nonrotating hole" {\it Phys. Rev.} {\bf 13}  2 198
\newline 
\noindent  
Custodio P S and  Horvath J E   2002 ``The evolution of primordial black hole masses in the radiation-dominated era"  {\it General Relativity and Gravitation} {\bf 34} 1895-1907 

\bibitem{visser} Barcelo C,  Liberati  S and Visser M 2011 ``Analogue gravity" {\it Living Reviews in Relativity} {\bf 14} 1-159 

\bibitem{Sandvik} Sandvik A  2010 ``Computational studies of quantum spin systems" {\it AIP Conference Proceedings} {\bf 1297} 1 135-338, arxiv:1101.3281v1

\bibitem{Kooning}  Kooning S and  Down M 1990 {\it Computational Physics. Fortran version} (Westview Press)  

\bibitem{ghezzi} Ghezzi C R 2017 ``Simulación numérica de superfluidos con autómata celulares" {\it Mecánica Computacional} {\bf 35} 25  1341-1350 

\bibitem{minnhagen} Weber H and Minnhagen P 1988 ``Monte Carlo determination of the critical temperature for the two-dimensional XY model" {\it Phys. Rev. B} {\bf 37} 10 5986 

\bibitem{bardeen} Bardeen J M,  Carter B and  S W Hawking 1973 ``The
four laws of black hole mechanics" {\it Commun. Math. Phys.} {\bf 31} 161 

\bibitem{susskind} Susskind L and Lindesay J 2005 {\it An introduction to black holes, information and the string theory revolution: The holographic universe} (World Scientific)

\bibitem{gibbons} Gibbons G W and  S W Hawking 1977  ``Cosmological event horizons, thermodynamics, and particle creation" {\it Phys. Rev. D} {\bf 15} 2738 

\bibitem{wells} Wells F S, {\it et al.} 2015 ``Analysis of low-field isotropic vortex glass containing vortex groups in YBa$_2$Cu$_3$O$_{7-x}$ thin films visualized by scanning SQUID microscopy" {\it Scientific Reports} {\bf 5} 8677 

\bibitem{Spisak} {Spi\v{s}\'{a}k D 1993 ``Self-consistent mean-field study of the anisotropic two-dimensional XY model" {\it Physica B: Condensed Matter} {\bf 190} 4 407-412}

\bibitem{frolov} Novikov I and Frolov V 2013 {\it Physics of black holes} (Springer Science \& Business Media)

\bibitem{ruffini} Damour T and Ruffini R 1976 ``Black-hole evaporation in the Klein-Sauter-Heisenberg-Euler formalism" {\it Phys. Rev. D} {\bf 14} 2 332 

\bibitem{huang2} Huang K, Xiong C and  Zhao X  2014 ``Scalar-field theory of dark matter'' {\it Int. J. Mod. Phys. A} {\bf 29} 13 1450074 

\bibitem{letelier} Lemos J  P and Letelier P S 1994  ``Exact general relativistic thin disks around black holes" {\it Phys. Rev. D} {\bf 49} 10 5135

\bibitem{ghezzi2} Ghezzi  C R  2005 ``Relativistic structure, stability, and gravitational collapse of charged neutron stars" {\it Phys. Rev. D} {\bf 72} 10 104017
\newline \noindent
 Ghezzi C R and Letelier P S 2007 ``Numeric simulation of relativistic stellar core collapse and the formation of Reissner-Nordström black holes" {\it Phys. Rev. D} {\bf 75} 2 024020 



\end{thebibliography}
\end{document}